# On the relation between mass of a pion, fundamental physical constants and cosmological parameters

Dragan Slavkov Hajdukovic[1]
PH Division CERN
CH-1211 Geneva 23
dragan.hajdukovic@cern.ch
[1]On leave from Cetinje, Montenegro

**Abstract**
In this article we reconsider the old mysterious relation, advocated by Dirac and Weinberg, between the mass of the pion, the fundamental physical constants, and the Hubble parameter. By introducing the cosmological density parameters, we show how the corresponding equation may be written in a form that is invariant with respect to the expansion of the Universe and without invoking a varying gravitational "constant", as was originally proposed by Dirac. It is suggested that, through this relation, Nature may give a hint on that virtual pions may dominate the "content" of the quantum vacuum.

Since the times of Dirac [1], a number of apparent numerical "coincidences" has attracted the attention of the physicists. As pointed out by Weinberg [2], the most striking numerical "coincidence" is, presumably, the relation

$$m_\pi^3 \sim \frac{\hbar^2}{cG} H_0 \tag{1}$$

which shows that the mass of a typical elementary particle, such as the pion, does not too much differ from a mass that is "constructed" from entirely out of the fundamental constants $G, \hbar, c$ and the Hubble constant $H_0$ (i.e. the present day value of the Hubble parameter $H$ ). It is not known whether Eq. (1) is just a (somewhat surprising) coincidence or whether it has any real physical significance. The attempt of writing the proportionality (1) as an equation encounters two problems: First, the left-hand side of Eq. (1) is about 10 times larger than the right-hand side; hence, in the case of an equality the "missing" proportionality factor on the right-hand side should come up for this order of magnitude discrepancy. The second problem is of more fundamental nature. In contemporary cosmology $H$ is not a constant; it is a function of the age of the universe. If we want the mass $m_\pi$ in (1) to be a constant, we must deal with the problem of a variable $H$. As for a possible solution Dirac [1] suggested that the ratio $H/G$ does not vary with cosmological time. This implied that $G$ would also vary with time (it would, in fact, decrease). According to Dirac, the reason why the gravitational force is so weak today is thus, because the universe is so old. This would be a beautiful theoretical explanation of the weakness of gravity. Unfortunately, Dirac's hypothesis is in conflict with observation; in particular, it provides a wrong age of the universe [2].

In the present letter an alternative approach is suggested.

Relation (1) contains only one "isolated" cosmological parameter (the Hubble parameter $H$ ). In the cosmological equations, $H$ always appears in combination with other cosmological parameters. For instance, in the case of a closed universe, the Friedmann equation is written as

$$\frac{c^2}{R^2 H^2} = \Omega - 1 \tag{2}$$

where $\Omega$ is the total energy density of the universe relative to the critical density and $R$ is the scale factor in the Friedmann-Robertson-Walker metric. Hence, in cosmology, $H$ , $R$ and $\Omega$ are inseparable, while in Eq. (1) $\Omega$ and $R$ do not appear. The right-hand side of Eq. (1) is, hence, *incomplete*, and $H_0$ should be replaced there with a quantity which is about 10 times larger, is independent of time, and in addition contains the density parameters $\Omega$ .

The same can be argued from a different point of view: If Eq. (1) would not be a mere coincidence, it suggests that pions possess some (hidden) importance for the Universe as a whole. How could this be possible? Pions are just a tiny fraction of the matter-energy in our world with quarks and leptons as its building blocks. If we insist on the validity of Eq. (1), one should "find" enormous quantities of pions somewhere in the Universe. Since large quantities of "real" pions do obviously not exist, it seems natural to suppose that the importance of pions Eq. (1) in this case suggests, should be due to "virtual" pions, which according to Quantum Field Theory are an inherent part of vacuum fluctuations. Yet, in the vacuum, there are other virtual particle-antiparticle pairs (for instance, electron-positron pairs). Eq. (1) may eventually suggest that pions in the vacuum are the dominant virtual pairs. If this would be so, the vacuum energy density parameter should somehow be included in Eq. (1).

In contemporary cosmology the "mass-energy content" of the Universe is successfully modelled as a perfect "cosmological fluid", which consists of a mixture of several distinct components (hereafter denoted by the subscript $n$ ). Each component of the cosmological fluid obeys an equation of state (i.e. relation between its pressure $p$ and its density $\rho$ ) of the form

$$p_n = w_n \rho_n c^2 \tag{3}$$

with constant equation-of-state parameter $w_n$ . Eq. (3) together with the power-law

$$\rho_n = \rho_{n,0} \left( \frac{R_0}{R} \right)^n \tag{4}$$

yields $w_n = (n-3)/3$. The index 0 in the above expression denotes the present-day values.

With the densities $\rho_n$ following the power-law (4) the corresponding density parameters $\Omega_n$ , defined as

$$\Omega_n = \frac{\rho_n}{\rho_{crit}} ; \;\; \rho_{crit} = \frac{3H^2}{8\pi G} \tag{5}$$

become

$$\Omega_n = \Omega_{n,0} \left( \frac{H_0}{H} \right)^2 \left( \frac{R_0}{R} \right)^n \tag{6}$$

In the Standard Cosmology, the cases $n = 4,3,0$ (with $w = 1/3, 0, -1$) correspond respectively to relativistic particles, pressureless matter (dust) and a cosmological constant $\Lambda$ (as one plausible candidate for dark energy); while $n = 2$ describes the time evolution of $(\Omega - 1)$. We may thus change the notation to $\Omega_4 = \Omega_r, \Omega_3 = \Omega_m, \Omega_2 = (\Omega - 1)$ and $\Omega_0 = \Omega_\Lambda$ .

Using Eq. (6), it is easy to show that the right-hand side of

$$m_\pi^3 = \frac{\hbar^2}{cG} H \left\{ \frac{\Omega_\Lambda}{\sqrt{\Omega - 1}} \frac{R_0}{R} \right\} \tag{7}$$

does not anymore depend on time, and there is no need to invoke a varying gravitational "constant". The factor that completes Dirac's relation (1) is put in brackets. It assures that Eq. (7) does not change during the expansion of the Universe, and it provides a numerical value of the order of about one order of magnitude. It may thus be considered as a solution of both problems that Dirac's original relation (1) had. It also supports the view that the "incomplete" Eq. (1) and its completed version (7) are more than numerical coincidences, but have some physical significance.

More generally, we may drop the assumption that the vacuum energy density is constant and denote the density parameter of the vacuum by $\Omega_v$ (replacing $\Omega_\Lambda$ ). As for an example, let's take $n = 1; \, w = -2/3$ . In this case we have

$$m_\pi^3 = \frac{\hbar^2}{cG} H \left\{ \frac{\Omega_v}{\sqrt{\Omega - 1}} \right\} \tag{8}$$

and expression that is even more elegant than Eq. (7). Of course, since, $\Omega_{\Lambda,0} = \Omega_{v,0}$, Eqs. (7) and (8) give the same result for the mass of the pion, independently of the future evolution of the vacuum density parameter. Eq. (8) illustrates how robust the Dirac relation (1) is and that it can be completed for different equations of state (3).

The mass of the pion and the fundamental constants are known much better than the cosmological parameters. We therefore rewrite Eq. (8) in the form

$$\frac{H\Omega_v}{\sqrt{\Omega-1}} = \frac{cG}{\hbar^2} m_\pi^3 = 2.507\times10^{-17}\, s^{-1} \tag{9}$$

The above numerical value (calculated by using the mass of the neutral pion) is in good agreement with the numerical values for $H_0, \Omega_0$ and $\Omega_{v,0}$, determined from observational cosmology [3].

The Eqs. (7) and (8) allow to write

$$m_\pi^3 = M_P^2 m_x \tag{10}$$

where $M_P = \sqrt{\hbar c/G}$ is the Plank mass, and

$$m_x = \frac{m_\pi^3}{M_P^2} = \frac{\hbar}{c^2}\frac{H\Omega_v}{\sqrt{\Omega-1}} = 2.93\times10^{-68}\, kg \tag{11}$$

The meaning of the mass $m_x$ is not evident, but let us point out two amusing comparisons: First, it was recently argued that in the framework of classical general relativity the presence of a positive cosmological constant implies the existence of a minimum mass in nature [4]; the above determined mass $m_x$ is just about one order of magnitude larger than the supposed minimum mass. Second, the mass $m_x$ is nearly identical to a recently conjectured mass of the graviton [5].

Another striking observation is that the mass of the neutrino is close to the geometrical mean of the Plank mass $M_P$ and the mass $m_x$:

$$m_\nu = \sqrt{M_P m_x} = 2.53\times10^{-38}\, kg \tag{12}$$

With the help of this relation, Eq. (10) can be rewritten in the form

$$m_\pi^3 = M_P m_\nu^2 \tag{13}$$

Moreover, the Plank mass is close to the geometrical mean of the mass $m_x$ and the mass of the Universe ($M_U$ ),

$$M_P \sim \sqrt{m_x M_U} \tag{14}$$

It is hard to believe that the relations (7) and (8), with their mathematical simplicity and structure that is based entirely on fundamental constants and cosmological parameters, are just numerical coincidences. It seems more plausible that they have some deeper rooted physical significance. In Quantum Chromodynamics (QCD), virtual pions represent an inherent part of the physical vacuum. In the framework of the Standard Model of elementary particle physics, the vacuum energy density receives contributions from any quantum field. Hence, QCD fields provide just one (and not the largest) contribution to the vacuum energy. The relations (7) and (8) suggest that, in contrast, only the QCD vacuum might somehow be important in Cosmology. They even seem to suggest that the matter-energy content of the Universe and the physical vacuum are so deeply interconnected that they cannot be studied separately.